\documentclass[aps,prl,twocolumn,floatfix,superscriptaddress]{revtex4}

\usepackage{amsmath}
\usepackage{amsfonts}
\usepackage{amssymb}
\usepackage{mathrsfs}
\usepackage{graphicx}
\usepackage{color}

\usepackage{subfigure}

\newcommand{\commutator}[2]{[#1,#2]}
\newcommand{\average}[1]{\langle#1\rangle}

\begin{document}

%\title{Force on a slow moving impurity due to thermal and 
%quantum fluctuations in a quasi-1D Bose-Einstein condensate}
\title{Drag force on an impurity below the superfluid critical velocity in a 
 quasi-one-dimensional Bose-Einstein condensate}
\author{Andrew G. Sykes}
\author{Matthew J. Davis}
\affiliation{The University of Queensland, School of Mathematics and Physics, 
ARC Centre of Excellence for Quantum-Atom Optics, Qld 4072, Australia.}
\author{David C. Roberts}
\affiliation{
Theoretical Division and Center for Nonlinear Studies, 
Los Alamos National Laboratory, Los Alamos, New Mexico 87545, USA}

\begin{abstract}
The existence of frictionless flow below a critical velocity for obstacles moving in a superfluid is well established in the
context of the mean-field Gross-Pitaevskii theory.  We calculate the next order correction due to 
quantum and thermal fluctuations and find a 
non-zero force acting on a delta-function impurity 
moving through a quasi-one-dimensional Bose-Einstein condensate at all subcritical 
velocities and at all temperatures.  The force occurs due to an imbalance in 
the Doppler shifts of reflected quantum fluctuations from either side of the impurity.
Our calculation is based on a consistent extension of Bogoliubov theory to second 
order in the interaction strength, and finds new analytical solutions to the Bogoliubov-de Gennes 
equations for a gray soliton. Our results raise questions regarding the quantum 
dynamics in the formation of persistent currents in superfluids.

\end{abstract}

\maketitle
Quantum fluids and their macroscopic manifestations have long intrigued 
physicists. 
Recent rapid advances in the experimental manipulation 
of dilute ultra cold gases~\cite{gorlitz,tolra,weiss,
hofferberth,amerongen,superfluid_expts} have opened a new window to the
understanding of macroscopic quantum phenomena. 
Such experiments provide an exciting opportunity to advance our theoretical 
understanding of quantum fluids as they probe regimes where calculations are more 
tractable than many traditional condensed matter systems.

One of the most remarkable features 
is the emergence of superfluidity and the resulting 
critical velocity, below which an obstacle is able to 
move without experiencing any friction~\cite{landau_1941}. 
The presence of a Bose-Einstein condensate (BEC) is commonly believed to be 
intrinsically related to superfluidity. 
In this paper we probe the superfluidity of a quasi-1D BEC with 
a delta-function 
impurity moving through it at constant velocity. 
We introduce a complete set of 
zero eigenvalue solutions to the Bogoliubov-de Gennes equations 
for the excitation spectrum 
of the BEC and impurity. Our calculation is 
consistent to second order in the interaction strength. 
We find a drag force is present at \emph{subcritical} 
speeds for all temperatures. 
We emphasize that this result is not in conflict with Landau's 
argument, as the drag force in this situation arises from 
the scattering of fluctuations 
(either quantum or thermal)  rather than the creation of 
quasiparticles~\cite{landau_1941}. 
This mechanism was first suggested as a source of dissipation at zero 
temperature via 
a perturbative calculation for a uniform three-dimensional BEC 
using the Born approximation~\cite{roberts1}. 
The current work goes beyond the 
perturbative regime and is applicable to repulsive 
delta-function impurities of any strength. 
Furthermore we calculate the force for systems at finite temperature,
 and find a clear distinction 
between the quantum and thermal regime. These results 
could be tested experimentally using foreseeable technology~\cite{kohl_preprint}.

The second-quantised Hamiltonian of the system 
in the frame moving with the impurity at velocity $v$ is
\begin{equation}
\hat{H}=
\int\!dx\,\,\hat{\psi}^\dagger\left(-\frac{\hbar^2}{2m}\partial_x^2+
i\hbar v\partial_x+\eta\delta(x)+
\frac{g}{2}\hat{\psi}^\dagger\hat{\psi}\right)\hat{\psi},
\label{eq:hamiltonian}
\end{equation}
where $\hat{\psi}=\hat{\psi}(x,t)$ is the bosonic field operator 
obeying the usual equal-time commutation relations, and $g>0$ 
is related to the 3D $s$-wave scattering length $a$: 
$g\simeq2\hbar\omega_\perp a$ (we have assumed a sufficiently tight 
transverse trapping potential of frequency $\omega_\perp$~\cite{olshanii}). 
%, 
%$\commutator{\hat{\psi}(x,t)}{\hat{\psi}^\dagger(x',t)}=\delta(x-x')$ and 
%$\commutator{\hat{\psi}(x,t)}{\hat{\psi}(x',t)}=\commutator{\hat{\psi}
%^\dagger(x,t)}{\hat{\psi}^\dagger(x',t)}=0$. 
We use the following dimensionless coordinates for the system. 
Time: $\tau=(gn_\infty/\hbar) t$, 
and length: $z=x/\xi$ where $\xi=\hbar/\sqrt{mgn_\infty}$ is the 
healing length, and $n_\infty$ 
is the total density, at $x=\pm\infty$ 
(far away from the impurity). 
The impurity strength and impurity velocity are 
parameterised by 
$\bar{\eta}=\eta/(v_s\hbar)>0$ and $\bar{v}=v/v_s$ respectively,
where $v_s=\sqrt{gn_\infty/m}$ is the speed of sound far from the impurity. 
Finally we rescale the field operator, 
$\hat{\psi}=\sqrt{n_\infty}\hat{\varphi}$. 
In these units the commutation relations become 
$\commutator{\hat{\varphi}(z,\tau)}{\hat{\varphi}^\dagger(z',\tau)}
=\sqrt{\gamma}\delta(z-z')$
where $\gamma=mg/(\hbar^2n_\infty)$ is the ratio of interaction 
energy to kinetic energy. %Although this far from constitutes 
%a proof, one can see that as 
%$\sqrt{\gamma}\rightarrow0$ the field operator behaves more like a 
%classical field which commutes. 
Our calculations are based on the 
assumption that $\epsilon\equiv\gamma^{1/4}\ll1$ and this defines 
the small parameter in our perturbative expansion.
The equation of motion for $\hat{\varphi}$ is then
\begin{eqnarray}
i\partial_\tau\hat{\varphi}%&=&\commutator{\hat{\varphi}}{\hat{H}}\nonumber\\
&=&\left(-\frac{1}{2}\partial_z^2+
i\bar{v}\partial_z+\bar{\eta}\delta(z)+
\hat{\varphi}^\dagger\hat{\varphi}\right)\hat{\varphi}.
\label{eq:op_motion}
\end{eqnarray}

An understanding of the relevant length scales 
in a 1D Bose gas is important~\cite{petrov}. 
The ground state of an \emph{infinitely} extended, weakly interacting 
1D Bose gas is that of a quasicondensate~\cite{quasicondensate}, 
with phase coherence length: 
$l_\phi=\xi\exp(\sqrt{{\pi^2}/{\gamma}})$. 
Hence phase coherence is 
present across regions of size $l_\phi\ggg\xi$, but not over all 
space. %(as is the case with a true condensate). 
We define the size of our system, $L$, to be the 
same order as the phase coherence length. 
That is $L\ggg\xi$ and $L\lesssim l_\phi$, such that we have a 
true condensate which can be considered 
homogeneous \footnote{The single particle spatial
correlation function decays as a power law at $T=0$, and exponentially 
for $T\neq0$. However, we consider small enough temperatures that the 
phase coherence length remains much larger than the healing length.}. 
We emphasize that we are not considering regimes of 1D systems for 
which there does not exist a true-BEC~\cite{pitaevskii_astrakharchik}.

We ultimately wish to calculate the force on the impurity, given by the 
gradient of the potential, 
$F=-\average{\hat{\varphi}^\dagger\partial_z\left[\bar{\eta}\delta(z)\right]\hat{\varphi}}$. 
First we find the solution to Eq. \eqref{eq:op_motion} from 
Bogoliubov's perturbation expansion  
$\hat{\varphi}\simeq(\varphi_0+\epsilon\hat{\varphi}_1+\epsilon^2\varphi_2)
e^{-it(\mu_0+\epsilon\mu_1+\epsilon^2\mu_2)}$. 
The expansion is 
conceptually summarised as 
(i) $\varphi_0$ is the condensate wave function in the absence of all 
fluctuations; (ii) $\hat{\varphi}_1$ describes the fluctuations 
of the field in the presence of a condensate
$\varphi_0$; 
(iii) $\varphi_2$ is the modification of the condensate 
wave function due to the presence of fluctuations. 
%We calculate the Bogoliubov expansion to second 
%order in $\epsilon$ where changes in the density occur. 

\begin{figure}
\includegraphics[width=0.4\textwidth]{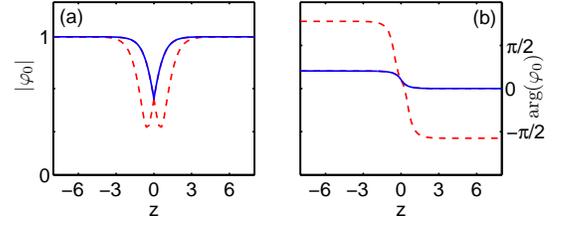}
\caption{(color online). Mean field solutions for (a) the condensate 
amplitude $|\varphi_0(z)|$ and (b) the condensate phase from
Eq.~\eqref{eq:psi0} for 
$\bar{\eta}=1$ and $v/v_s=0.35$. 
The dashed line shows the continuation of each of the 
underlying dark soliton solutions. 
}
\label{fig:gpe}
\end{figure}

(i) {$\mathcal{O}(\epsilon^0)$:} At zeroth order the commutators 
vanish and 
%we simply use 
a $c$-number field describes the system. 
The condensate wave function
%: $\hat{\varphi}\rightarrow\varphi_0e^{-it}$ 
is given by the solution to the Gross-Pitaevskii equation 
\begin{equation}
L[\varphi_0]\varphi_0=0,\label{eq:gpe}
\end{equation}
where $L[\varphi_0]\equiv-\frac{1}{2}\partial_z^2+i\bar{v}\partial_z
+\bar{\eta}\delta(z)+|\varphi_0|^2-1$ ($\mu_0=1$).
This can be 
solved analytically~\cite{hakim} 
for impurity velocities less than the critical velocity 
$\bar{v}<\bar{v}_c$, where $\bar{v}_c$ depends on $\bar{\eta}$ 
by $\bar{v}_c^2=1-\frac{2\bar{\eta}^2}{3}+
\frac{\bar{\eta}}{\sqrt{2}}\left(\frac{3}{R}-R\right)+\bar{\eta}^2\left(
\frac{R^2}{9}+\frac{1}{R^2}\right)$ where 
$R^3\equiv2\sqrt{2}\bar{\eta}/\left(\sqrt{1+8\bar{\eta}^2/27}-1\right)$~\cite{gunn}. 
%One first replaces 
To do so, the term 
$\bar{\eta}\delta(z)$ in Eq.~\eqref{eq:gpe} is replaced with a \emph{derivative jump}: 
$\partial_z\varphi_0(z=0^+)-\partial_z\varphi_0(z=0^-)=2\bar{\eta}$. 
The dark soliton solutions~\cite{tsuzuki} are used to give
\begin{equation}
\varphi_0(z)=\left\{
\begin{array}{ll}
e^{i\sigma_-}\left\{\cos(\theta)
\tanh\left[z_c^-\right]
+i\sin(\theta)\right\} & (z<0),\\
e^{i\sigma_+}\left\{\cos(\theta)
\tanh\left[z_c^+\right]
+i\sin(\theta)\right\} & (z>0),
\end{array}\right.\label{eq:psi0}
\end{equation}
where the impurity velocity is parameterised by $\bar{v}=\sin(\theta)<\bar{v}_c$, 
and $z_c^\pm=(z\pm z_0)\cos(\theta)$. The phases $\sigma_+=-\theta$, and 
$\sigma_-=\theta-\pi+2\arctan[\sin(2\theta)/(\exp(2z_0\cos\theta)-\cos(2\theta))]$ 
are fixed, although $\varphi_0$ has an arbitrary global phase 
factor. %Due respect is given to the authors of Refs~\cite{castin_dum_ggpe,morganggpe} 
%for calculating the effects of symmetry breaking assumptions. 
Here, we work in the limit of a large system, where the 
effects of fluctuations in the total 
number of particles are negligible~\cite{castin_dum_ggpe}. 
The quantity $z_0$ is determined by the derivative 
jump condition, which 
reduces to numerically solving $\bar{\eta}=\cos^{3}(\theta)
\tanh\left[\cos(\theta)z_0\right]/
\left(\sin^2(\theta)+\sinh^2\left[\cos(\theta)z_0\right]\right)$ for a given value 
of $\bar{\eta}$ and $\bar{v}<\bar{v}_c$. Two solutions exist 
for $z_0$ but only one is physical~\cite{hakim}. Figure \ref{fig:gpe} 
shows a plot of $\varphi_0(z)$ for a specific $\bar{\eta}$ and 
$\bar{v}$. As expected, at this level of approximation the drag force $F=0$. For numerical solutions to 
Eq. \eqref{eq:gpe} in higher dimensions see Ref.~\cite{2d_gpe}.

(ii) {$\mathcal{O}(\epsilon^1)$:} 
The quantum and thermal depletion of the
condensate is accounted for at first order. %The quantity 
%$\hat{\varphi}_1$ is the first order approximation to the 
%fluctuations of $\hat{\varphi}$ in the sense that $\average{\hat{\varphi}}=\varphi_0$ 
%and $\average{\hat{\varphi}_1}=0$ (technically we are taking the average in a restricted 
%ensemble where the phase symmetry has been broken). 
Linearisation of 
Eq. \eqref{eq:op_motion} with respect to $\epsilon$ results in 
\begin{equation}
i\partial_\tau\hat{\varphi}_1=
%\left[-\frac{1}{2}\partial_z^2+i\bar{v}\partial_z
%+\bar{\eta}\delta(z)+2|\varphi_0|^2-1\right]\hat{\varphi}_1+
%\varphi_0^2\hat{\varphi}_1^\dagger
L[\sqrt{2}\varphi_0]\hat{\varphi}_1+\varphi_0^2\varphi_1^\dagger.
\label{eq:quant_fluct_eq_motion}
\end{equation}
The density is unaffected at order $\epsilon$ (as $\langle\hat{\varphi}_1\rangle=0$) and therefore 
$\mu_1=0$. 
A Bogoliubov transformation is then performed, 
$\hat{\varphi}_1=\int\!dk\left[
u_k(z)e^{-iE_k\tau}\hat{\alpha}_k+
v_k^*(z)e^{iE_k\tau}\hat{\alpha}_k^\dagger\right]
$ 
where $\hat{\alpha}_k$ ($\hat{\alpha}_k^\dagger$) are the 
annihilation (creation) operators of the 
excitations (note that $k$ is a continuous index). 
They obey the usual commutation relations, 
$[\hat{\alpha}_k,\hat{\alpha}_{k'}^\dagger]=\delta(k-k')$ and 
$[\hat{\alpha}_k,\hat{\alpha}_{k'}]=
[\hat{\alpha}^\dagger_k,\hat{\alpha}^\dagger_{k'}]=0$. 
The quantum state of 
the system is defined such that the occupation number of each excitation is 
$n_k\equiv\average{\hat{\alpha}_k^\dagger\hat{\alpha}_k}=1/(e^{E_k/T}-1)$. 
%In this way we are making the assumption that one can get far enough 
%away from the impurity, such that the system has no knowledge of the impurity. 
%This is an inherently non-equilibrium assumption in spite of the 
%remarkable fact that, at 
%the level of the condensate wave function, 
%this assumption is completely consistent with equilibrium.
The amplitudes, $u_k$ and $v_k$, 
are determined by
\begin{equation}
\left[\begin{array}{cc}L[\sqrt{2}\varphi_0] & \varphi_0^2 \\
-(\varphi_0^*)^2 & 
-L^*[\sqrt{2}\varphi_0]
\end{array}
\right]
%\left[\begin{array}{c}
%u_k\\
%v_k
%\end{array}
%\right]
\vec{u}_k=E_k
%\left[\begin{array}{c}
%u_k\\v_k\end{array}
%\right]
\vec{u}_k,\label{eq:bdg_eqs}
\end{equation}
%known as the Bogoliubov-deGennes equation. 
where $\vec{u}_k=(u_k,v_k)^\textrm{T}$. 
The normalisation which preserves the bosonic 
commutation relations is given by 
$\int dz\vec{u}_k^{\:\dagger}\sigma_z\vec{u}_{k'}=\delta(k-k')$ where
$\sigma_z$ is a Pauli matrix.

The solutions of Eq.~\eqref{eq:bdg_eqs} for $z\in\mathbb{R}^\pm$ 
are~\cite{squared_jost_solns,bilas_pavloff_darksoliton}
\begin{equation}
%\left[\begin{array}{c}\zeta_k^\pm(z) \\
%\xi_k^\pm(z)\end{array}\right]
\vec{\xi}_k^\pm\equiv e^{ikz}
\left[\begin{array}{c}
e^{i\sigma_\pm}\left(\frac{k}{2}+
\frac{E_k}{k}+i\cos\theta\tanh\left(z_c^\pm\right)\right)^2 \\
e^{-i\sigma_\pm}\left(\frac{k}{2}-
\frac{E_k}{k}+i\cos\theta\tanh\left(z_c^\pm
\right)\right)^2\end{array}
\right],\label{eq:squaredjost}
\end{equation} 
where $E_k$ has two distinct branches: $E_k=E_k^\pm\equiv \pm k\left(
\mp\sin(\theta)+\sqrt{k^2/4+1}\right)$, % or 
%equally $E_k=E_k^\leftarrow\equiv -k\left(\sin(\theta)+\sqrt{k^2/4+1}\right)$. 
corresponding to 
the left and right propagating excitations 
and hence $E_k>0$ (special attention 
must be paid to the $E_k=0$ mode). For a 
fixed excitation energy $E$, there are four possible 
wavenumbers, given by the four roots of  
$(1/4)k^4+\cos^2(\theta)k^2+2E\sin(\theta)k-E^2=0$. Two of the roots will be in 
$\mathbb{R}$, denoted $k$ and $k_r$ with one positive and one negative. 
$k_r$ is a \emph{reflected} mode, and $k_r=-k$ \emph{only} 
when $v=0$. For $v\neq0$ a Doppler shift will occur. 
The other two roots will be complex conjugates, (with nonzero imaginary parts) denoted 
$k_e$ and $k_e^*$. 
%These modes exponentially 
%increase/decrease in space. 
On either side of the impurity, one must select the 
exponentially \emph{decreasing} mode to satisfy the boundary
conditions of the problem. 
%to yield the evanescent component of the 
%scattering process. 
%Exponentially increasing modes are prohibited, to satisfy the boundary conditions. 
Each wavenumber corresponds to a particular solution of Eq.~\eqref{eq:bdg_eqs},
which is 
ultimately a 4th order differential equation. 
The normalisation of the eigenvectors in Eq.~\eqref{eq:squaredjost} is 
%${N_k^2}\int dz\left[(\zeta_k^\pm)^*\zeta_{k'}^\pm-(\xi_k^\pm)^*\xi_{k'}^\pm\right]=
%\delta(k-k')$ 
${N_k^2}\int dz\vec{\xi}_k^{\:\dagger}\sigma_z\vec{\xi}_k=
\delta(k-k')$ where $N_k=\frac{1}{4}\sqrt{k/\pi}(k^2/4+1)^{-1/4}E_k^{-1}$. 
%It can be seen that with this normalisation, the 
Thus the 
solutions in Eq.~\eqref{eq:squaredjost} 
correspond exactly to the 
usual Bogoliubov excitations of a uniform BEC traveling at velocity $v$ 
for $z\rightarrow\pm\infty$~\cite{fetter2}. 

To incorporate the impurity at $z=0$, we separate the problem into two 
independent scattering problems: One in which the incoming mode approaches from $z=-\infty$ and 
the other in which it approaches from $z=+\infty$. 
%The modes generated 
%by the two separate problems 
%do not interfere, since there is no well defined phase between them. 
%In this scattering problem, 
The boundary conditions we impose are: (i) the amplitude 
of the incoming wave is given by $N_k$ in accordance with the 
assumption that, away from the impurity, the BEC has no knowledge of 
the impurity. Implicit in this condition is a time scale 
over which it is assumed the scattered waves have not reached 
edge of the system. 
(ii) Scattered waves cannot exponentially increase as they move 
away from the impurity. (iii) Scattered waves must be \emph{causally} related 
to the incoming wave. %These is an inherently non-equilibrium assumption in spite 
%of the remarkable fact that, at the level of the condensate wave function it is 
%completely consistent with equilibrium. 
For the case where the 
incoming wave approaches from $z=-\infty$ the solution is
\begin{equation}
\vec{u}_k=\left\{\begin{array}{ll}
N_k\vec{\xi}^-_k+A_r\vec{\xi}^-_{k_r}+A_{er}\vec{\xi}^-_{k_e^*} & (z<0),\\
A_t\vec{\xi}^+_k+A_{et}\vec{\xi}_{k_e}^+ & (z>0),
\end{array}\right.
%\left[\begin{array}{c}
%u_k\\
%v_k\end{array}
%\right]=
%\left\{
%\begin{array}{ll}
%N_k\left[\begin{array}{c}
%\zeta_k^-\\
%\xi_k^-\end{array}\right]
%+A_{r}\left[\begin{array}{c}
%\zeta_{k_{r}}^-\\
%\xi_{k_{r}}^-\end{array}\right]
%+A_{er}\left[\begin{array}{c}
%\zeta_{k_{e}^*}^-\\
%\xi_{k_{e}^*}^-\end{array}\right] &
%(z<0)
%\\
%A_{t}\left[\begin{array}{c}
%\zeta_{k}^+\\
%\xi_{k}^+\end{array}\right]
%5\zeta_{k_{e}}^+\\
%\xi_{k_{e}}^+\end{array}\right]&
%(z>0)
%\end{array}
%\right.
\label{eq:bdgsoln}
\end{equation} 
where $k>0$ and the complex constants $A_r$, $A_t$, $A_{et}$, and $A_{er}$ are 
completely determined by 
%\begin{align}
$\vec{u}_k|_{z=0^+}=\vec{u}_k|_{z=0^-}$ and
$\partial_z\vec{u}_k|_{z=0^+}-\partial_z
\vec{u}_k|_{z=0^-}=2\bar{\eta}\vec{u}_k|_{z=0},$ 
%\left[\begin{array}{c}
%u_k\\
%v_k\end{array}
%\right]_{z=0^-}&=
%\left[\begin{array}{c}
%u_k\\
%v_k\end{array}
%\right]_{z=0^+}\\
%\partial_z\left[\begin{array}{c}
%u_k\\
%v_k\end{array}
%\right]_{z=0^+}
%-\partial_z&
%\left[\begin{array}{c}
%u_k\\
%v_k\end{array}
%\right]_{z=0^-}=2\bar{\eta}
%\left[\begin{array}{c}
%u_k\\
%v_k\end{array}
%\right]_{z=0}
%%
%\begin{array}{c}u_k(z=0^-)=u_k(z=0^+) \\
%v_k(z=0^-)=v_k(z=0^+)\\
%\partial_zu_k(z=0^+)-\partial_zu_k(z=0^-)=2\bar{\eta}u_k(z=0)\\
%\partial_zv_k(z=0^+)-\partial_zv_k(z=0^-)=2\bar{\eta}v_k(z=0)
%\end{array}\nonumber
%\end{align}
which arise from the $\bar{\eta}\delta(z)$ term in Eq. \eqref{eq:bdg_eqs}. 
Solving the scattering problem where the incoming wave 
approaches from $z=+\infty$ follows the same logic, except now $k<0$ 
and the $z>0$ and $z<0$ conditions in Eq. \eqref{eq:bdgsoln} have to be 
swapped. 

At this stage the problem has been reduced 
to a set of linear algebraic equations 
for each value of $k$. Analytic progression is hindered by the fourth order 
polynomial determining $k_e$ and $k_r$. However, quantities such as: 
$\langle\hat{\varphi}_1^\dagger(z)\hat{\varphi}_1(z)\rangle=\int dk[|u_k|^2n_k+|v_k|^2(1+n_k)]$ and 
$\langle\hat{\varphi}_1(z)\hat{\varphi}_1(z)\rangle=\int dk[
u_kv_k^*(1+2n_k)]$ are easily attained numerically to very high 
accuracy once one fixes $\bar{\eta}$, $\bar{v}$, and $T$. 
One must include a 
small $k$ cut-off [of order $\sim1/L$] to prevent 
long wavelength fluctuations destroying the 
long range order in the system. Results are logarithmically dependent on 
the choice of this cut-off~\cite{russian1dnotes}.
%\begin{figure}
%includegraphics[width=0.4\textwidth]{scattering_picture_prl.eps}
%\caption{shows the scattering channels available to the 
%Bogoliubov modes. ... (I don't think this figure is appropriate).}
%\end{figure}

(iii) {$\mathcal{O}(\epsilon^2)$:} %These terms describe 
%the \emph{back-action} on the condensate wave function due to the 
%presence of $\hat{\varphi_1}$.
At second order, the generalised-Gross-Pitaevskii 
equation~\cite{castin_dum_ggpe} is
\begin{equation}
\mathcal{H}|\varphi_2(z)\rangle=|f(z)\rangle,\label{eq:ggpe}
\end{equation}
where  $f(z)=2\varphi_0\int dk
[|u_k|^2n_k+|v_k|^2(n_k+1)]+\varphi_0^*\int dk[u_kv_k^*(1+2n_k)]
-\mu_2\varphi_0$, (this definite integral is performed numerically using the 
analytic results of the previous section), and 
\begin{equation}
\mathcal{H}=\left[\begin{array}{cc}L[\sqrt{2}\varphi_0] & \varphi_0^2 \\
(\varphi_0^*)^2 & 
L^*[\sqrt{2}\varphi_0]
\end{array}
\right].
\end{equation}
%\begin{equation}
%L[\sqrt{2}\varphi_0]\varphi_2+\varphi_0^2\varphi_2^*=f(z)\label{eq:ggpe}
%\end{equation}
Here, a ket always corresponds to a complex conjugate pair: $|\bullet\rangle\equiv
(\bullet^{\phantom{*}},\bullet^*)^\textrm{T}$. 
The shift in the chemical potential, 
$\mu_2$ is defined on either side of 
the impurity such that $f(z)\rightarrow0$ as $z\rightarrow\pm\infty$, and contains 
the necessary terms to ensure orthogonality between the condensate mode and the 
excitations~\cite{castin_dum_ggpe}.

To solve Eq.~\eqref{eq:ggpe} for $\varphi_2(z)$, we use four linearly independent 
solutions of the homogeneous equation, 
$\mathcal{H}|\omega\rangle=|0\rangle$, 
to construct a $2\times2$ Green's matrix, $\mathcal{G}(z,s)$. 
Finding these four linearly independent solutions constitutes a 
nontrivial step, and we were unable to find 
expressions for them in the previous literature; we provide a 
detailed presentation of these solutions in~\cite{epaps}. 
Briefly, linear combinations of Eqs.~(2--5) of~\cite{epaps} 
are used to generate four solutions, denoted $|\Lambda_j(s)\rangle$, appropriate to the 
presence of an impurity [the $\bar{\eta}\delta(z)$ 
term in Eq.~\eqref{eq:ggpe}]. 
The Green's matrix satisfies
\begin{equation}
\mathcal{H}\mathcal{G}(z,s)=
\delta(z-s)\mathbb{I}_2,\label{eq:geq}
\end{equation}
where $\mathbb{I}_2$ is the $2\times2$ identity. 
This condition yields Eqs.~(13--16) of Ref.~\cite{epaps}. 
The boundary conditions for Eq. \eqref{eq:ggpe} are: 
(i) $\varphi_2$ does not exponentially increase away 
from the impurity; (ii) $\varphi_2$ is symmetric about the 
impurity for $v=0$; (iii) Any non-zero drag force 
on the impurity 
acts to decrease the relative motion between condensate 
and impurity. 
These conditions combined with the symmetry of the Green's matrix 
[specifically $\mathcal{G}(z,s)=\mathcal{G}^{\dagger}(s,z)$ which arises 
from the operator $\mathcal{H}$ being Hermitian] uniquely 
define the solution
\begin{equation}
|\varphi_2(z)\rangle=\int \!ds \;\mathcal{G}(z,s)
|f(s)\rangle,\label{eq:ggpesoln}
\end{equation}
up to a global phase factor. 
We note that Eq.~\eqref{eq:ggpesoln} is not a general formula for any function 
$f(s)$ --- 
the fact that $f(s)$ decays to zero as $s\rightarrow\pm\infty$ compensates 
for the linear divergences in $\mathcal{G}$. % as $s\rightarrow\pm\infty$. %Without 
%this decay, the solution could not be attained via this method. % Figure 
%\ref{fig:greensmatrix} shows a graph of $\mathcal{G}$. Note 
%the discontinuities in the derivatives due to both the impurity [along $z,s=0$], and the 
%conditions of Eq. \eqref{eq:geq} [along $z=s$]. 
The integral in Eq.~\eqref{eq:ggpesoln} is also performed numerically.%, 
%using the $f(z)$ obtained from the previous section.

%\begin{figure}
%\includegraphics[width=0.45\textwidth]{green_graph.eps}
%\caption{shows the real and imaginary of components of 
%$\mathcal{G}_{11}(z,s)$ and $\mathcal{G}_{12}(z,s)$.}
%\label{fig:greensmatrix}
%\end{figure}

\begin{figure}
\includegraphics[width=0.5\textwidth]{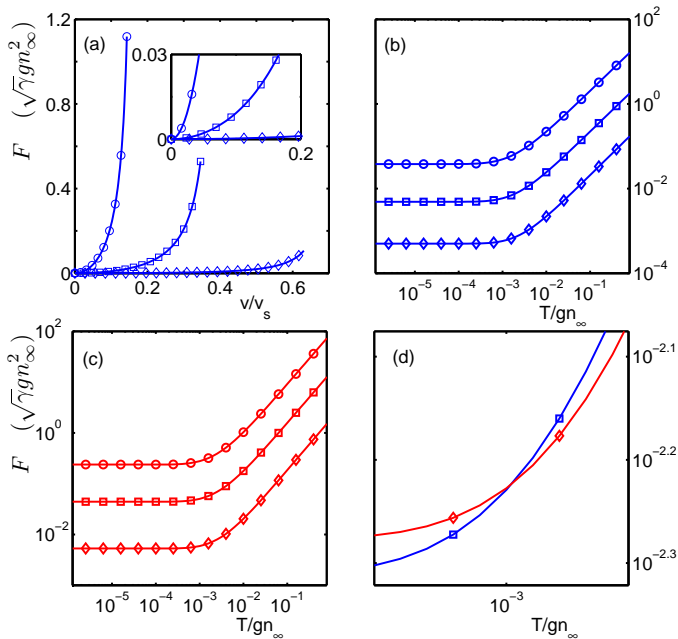}
\caption{(color online). The drag force on the impurity (in units $\sqrt{\gamma}gn_\infty^2$). 
(a) Drag force at $T=0$ as a function of velocity. $\bar{\eta}=0.3, 1, 3$ for diamond, square, circle respectively, and each line 
terminates at their respective critical velocities $v_c/v_s=0.64, 0.37, 0.16$. (b--d) Drag force as a 
function of temperature. (b) 
$\bar{\eta}=0.3$ and velocities: 
blue diamond $\bar{v}=0.14$, blue square $\bar{v}=0.35$, blue circle $\bar{v}=0.56$. 
(c) $\bar{\eta}=1$ and velocities: 
red diamond $\bar{v}=0.08$, red square: $\bar{v}=0.2$, and red circle $\bar{v}=0.3$. 
(d) is data taken from (b) and (c), and compares the drag on a strong impurity 
moving at slow speed with that on 
a weak impurity moving at high speed. The strong impurity can 
have a larger drag in the quantum regime while the weak impurity 
will have the larger drag in the thermal regime.}
\label{fig:forcegraph}
\end{figure}

%\begin{figure}
%\centering
%\subfiguretopcaptrue
%\subfigure[]{\label{fig:zerotemp}\includegraphics[width=0.4\textwidth]{zero_temp_force.eps}}
%\hspace{1cm}
%\subfigure[]{\label{fig:finitetemp}\includegraphics[width=0.4\textwidth]{finite_temp_force.eps}}
%\centering
%\caption{The drag force on the impurity at (a) zero 
%temperature ($\bar{\eta}=0.3, 1, 3$ for diamond, square, circle respectively) 
%and (b) finite temperature. In (b) the blue, red lines are for 
%$\bar{\eta}=0.3,1$ respectively. The velocities are: 
%blue diamond $\bar{v}=0.14$, blue square $\bar{v}=0.35$, blue circle $\bar{v}=0.56$, 
%red diamond $\bar{v}=0.08$, red square: $\bar{v}=0.2$, and red circle $\bar{v}=0.3$.}
%\label{fig:forcegraph}
%\end{figure}

We now evaluate the force to $\mathcal{O}(\epsilon^2)$
\begin{equation}
F\simeq\bar{\eta}\partial_z\left[\int dk[|u_k|^2n_k+|v_k|^2(1+n_k)]+
2\textrm{Re}\left\{\varphi_0\varphi_2^*\right\}\right],
\end{equation}
where all the derivatives are performed analytically using 
 Eq.~\eqref{eq:bdgsoln} 
and Eqs.~(2--5) of~\cite{epaps}. 
The results are shown in Fig.~\ref{fig:forcegraph}(a) 
for the case of zero temperature. 
The fact that the force is \emph{nonzero} below the critical 
velocity may be unexpected, although we note 
the magnitude is much smaller than the supercritical forces 
present above $v_c$~\cite{hakim}. We emphasise 
that quasiparticle creation is 
\emph{not} the mechanism behind this drag force, and consequentially this is not a 
contradiction of Landau's statements~\cite{landau_1941}. 
At zero temperature fluctuations in the quantum vacuum scatter from 
each side of the impurity, 
and the asymmetric Doppler shift in the reflected waves 
%from the two 
%separate scattering problems (incoming waves from $z=\pm\infty$) 
results in an imbalance and a non-zero net force.
A crucial assumption in this result is that the incoming waves 
have no knowledge of the impurity. 
In any finite system this assumption would \emph{only} be valid 
up to a time $t_c\sim L/v_s$ after the initial acceleration of the impurity. 
It could be imagined that the force on the impurity 
will decay to zero over this time scale 
$t_c$, which can be thought of as the relaxation 
 time of the quantum vacuum. 
We intend to  test this conjecture by performing dynamical simulations 
that incorporate the effects of the quantum fluctuations; this significant
project lies beyond the scope of this work.

The results at finite temperature are shown in Fig.~\ref{fig:forcegraph}(b--d).
The quantum and thermal regimes are clearly distinguished with a cross-over 
near $T\sim10^{-3}gn_\infty$. %This would correspond to 
%temperatures around $0.1$ nK in typical 
%cold atom experiments. 
The force increases linearly with $T$ in the thermal 
regime due to 
the scattering being most prominent for the 
low energy modes whose occupation is 
$n_k =1/(e^{E_k/T}-1)\sim T/E_k$ for $T\gg E_k$. The reflected modes for the high energy
incoming waves have a 
negligible Doppler shift and only a small contribution to the force.

In conclusion, we have calculated the drag force acting on an impurity 
moving through a true BEC, in a quasi-1D geometry, using 
analytic solutions of the Bogoliubov perturbation expansion. 
We find that a nonzero force arises for zero and finite temperature at 
all velocities due to an imbalance in the Doppler shift in the scattering 
of quantum and thermal fluctuations. 
The crossover from the quantum to the thermal regime occurs at temperatures near
 $T\sim10^{-3}gn_\infty$.
We have proposed a mechanism that predicts a time scale 
of $L/v_s$ for which it would be possible to observe the force in a finite system.

The authors wish to thank Tod Wright, Dmitry Petrov, Qian Niu, and Nicolas Pavloff 
for insightful discussions. AGS acknowledges the 
hospitality and support of 
the Center for Nonlinear Studies. AGS and MJD acknowledge support from the 
Australian Research Council Centre of Excellence for Quantum-Atom Optics.

\begin{widetext}

\section{AUXILLIARY PUBLICATION FOR: Drag force on an impurity below the superfluid
critical velocity in a quasi-one-dimensional Bose-Einstein condensate}

Here we describe the solution of Eq.~(11) in the main text (repeated here for
convenience)
\begin{equation}
\mathcal{H}\mathcal{G}(z,s)=
\delta(z-s)\mathbb{I}_2,\label{eq:geq1}
\end{equation}
where $\mathbb{I}_2$ is the $2\times2$ identity. 
We begin by writing down the four linear independent solutions to the homogenous
equation
\begin{equation}
\mathcal{H}|\omega\rangle=|0\rangle\label{eq:ggpehomo}
\end{equation}
for the separate cases of $z\in\mathbb{R}^+$ and $z\in\mathbb{R}^-$. 
These are denoted 
$|\omega_i^\pm(z)\rangle$ (recall that the ket notation refers to a complex conjugate pair) 
where, 
\begin{align}
\omega_1^\pm(z\in\mathbb{R}^\pm)&=i\varphi_0^\pm,\label{eq:om1}\\
\omega_2^\pm(z\in\mathbb{R}^\pm)&=e^{i\sigma_\pm}\textrm{sech}^2(z_c^\pm),\label{eq:om2}\\
\omega_3^\pm(z\in\mathbb{R}^\pm)&=e^{i\sigma_\pm}\left(\textrm{sech}^2(z_c^\pm)\left[2z_c^\pm-z_c^\pm\cosh(2z_c^\pm)
+(3/2)\sinh(2z_c^\pm)\right]\tan(\theta)+2i\left[z_c^\pm\tanh(z_c^\pm)-1
\right]\right),\label{eq:om3}\\
\omega_4^\pm(z\in\mathbb{R}^\pm)&=e^{i\sigma_\pm}\textrm{sech}^2(z_c^\pm)\left[z_c^\pm\left(10-4\cos^2(\theta)-
8\sin(\theta)\sin(\theta-2iz_c^\pm)\right)+\right.\nonumber\\
&\qquad\quad\left.
\cosh(z_c^\pm)\left[i\sin\left(2\theta-3iz_c^\pm\right)-5i\sin\left(2\theta
-iz_c^\pm\right)\right]+6\sinh(2z_c^\pm)\right].\label{eq:om4}
\end{align}
To the best of our knowledge, Eqs. \eqref{eq:om3} and \eqref{eq:om4} do not 
appear in the previous literature. 
Note from these solutions, the behaviour as $z\rightarrow\pm\infty$: 
$\omega_1^\pm\rightarrow$ constant, $\omega_2^\pm\rightarrow$ 0, 
$\omega_3^\pm\rightarrow$ linear increase/decrease, and 
$\omega_4^\pm\rightarrow$ exponential increase/decrease, which can be 
seen from Fig. \ref{fig:zeroeig}. 
\begin{figure}[b]
\includegraphics[width=0.9\textwidth]{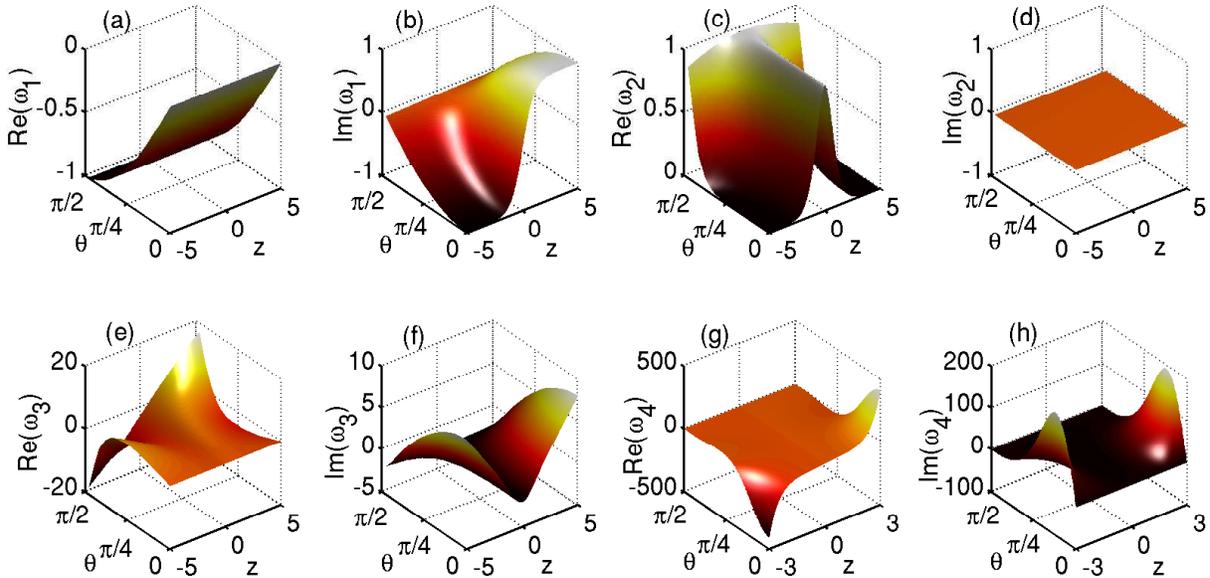}
\caption{(color online). The zero eigenvalue solutions of Eq.~\eqref{eq:ggpehomo} for
$z_0=0$ and $\sigma_\pm=0$ as a function of $z$ and $\theta$. 
(a) and (b) show the real and imaginary parts of $\omega_1(z)$, 
(c) and (d) show the real and imaginary parts of $\omega_2(z)$, 
(e) and (f) show the real and imaginary parts of $\omega_3(z)$,
and (g) and (h) show the real and imaginary parts of $\omega_4(z)$.}
\label{fig:zeroeig}
\end{figure}

Using these eight functions $|\omega_j^\pm(z)\rangle$ 
(where $j$ always runs over the indices $j=1,\ldots,4$) 
we construct four linearly independent solutions (denoted $|\Lambda_j(z)\rangle$) 
to Eq.~\eqref{eq:ggpehomo} for the entire domain $z\in\mathbb{R}$. 
This is done by 
matching the solutions in $\mathbb{R}^\pm$ across the point $z=0$ 
subject to the conditions 
\begin{subequations}
\begin{align}
|\Lambda_j(z=0^+)\rangle=&|\Lambda_j(z=0^-)\rangle,\\
\partial_z|\Lambda_j(z=0^+)\rangle-
\partial_z|\Lambda_j(z=&0^-)\rangle=2\bar{\eta}|\Lambda_j(z=0)\rangle,
\end{align}\label{eq:homobcs}
\end{subequations}
which are the usual conditions appropriate for 
the term $\bar{\eta}\delta(z)$ in the differential operator $\mathcal{H}$. 
The result is
\begin{align}
|\Lambda_1(z)\rangle&=\left\{\begin{array}{l}
|\omega_1^-(z)\rangle \\
|\omega_1^+(z)\rangle
\end{array}\right.\\
|\Lambda_2(z)\rangle&=\left\{\begin{array}{l}
|\omega_2^-(z)\rangle \\
a_1|\omega_1^+(z)\rangle+a_2|\omega_2^+(z)\rangle+
a_3|\omega_3^+(z)\rangle+a_4|\omega_4^+(z)\rangle
\end{array}\right.\\
|\Lambda_3(z)\rangle&=\left\{\begin{array}{l}
|\omega_3^-(z)\rangle \\
b_1|\omega_1^+(z)\rangle+b_2|\omega_2^+(z)\rangle+
b_3|\omega_3^+(z)\rangle+b_4|\omega_4^+(z)\rangle
\end{array}\right.\\
|\Lambda_4(z)\rangle&=\left\{\begin{array}{l}
|\omega_4^-(z)\rangle \\
c_1|\omega_1^+(z)\rangle+c_2|\omega_2^+(z)\rangle+
c_3|\omega_3^+(z)\rangle+c_4|\omega_4^+(z)\rangle
\end{array}\right.
\end{align}
where $a_j,b_j,c_j\in\mathbb{R}$ and are determined 
from Eqs.~\eqref{eq:homobcs} above.

The next step involves finding $\mathcal{G}(z,s)$ which is a solution 
to Eq. (11) of the main article. We write
\begin{equation}
\mathcal{G}(z,s)=\left\{\begin{array}{ll}
|\Lambda_1(z)\rangle\langle\lambda_1(s)|+
|\Lambda_2(z)\rangle\langle\lambda_2(s)|+
|\Lambda_3(z)\rangle\langle\lambda_3(s)|+
|\Lambda_4(z)\rangle\langle\lambda_4(s)|& \quad(z<s)
\\
&
\\
|\Lambda_1(z)\rangle\langle\kappa_1(s)|+
|\Lambda_2(z)\rangle\langle\kappa_2(s)|+
|\Lambda_3(z)\rangle\langle\kappa_3(s)|+
|\Lambda_4(z)\rangle\langle\kappa_4(s)|& \quad(z>s)
\end{array}\right.
\end{equation}
With some work it can be shown that the conditions
\begin{subequations}
\begin{align}
\mathcal{G}(z=s^+,s)=&\mathcal{G}(z=s^-,s)\\
\partial_z\mathcal{G}(z=s^+,s)-
\partial_z\mathcal{G}(z&=s^-,s)=-2\mathbb{I}_2
\end{align}
\end{subequations}
are completely equivalent to the following equations
\begin{align}
|\kappa_1(s)\rangle-|\lambda_1(s)\rangle&=\frac{1}{2}\sec^2(\theta)
|\Lambda_3(s)\rangle,\\
|\kappa_2(s)\rangle-|\lambda_2(s)\rangle&=\frac{1}{4}\sec(\theta)
\tan(\theta)|\Lambda_3(s)\rangle+\frac{1}{16}\sec^3(\theta)
|\Lambda_4(s)\rangle,\\
|\kappa_3(s)\rangle-|\lambda_3(s)\rangle&=-\frac{1}{2}\sec^2(\theta)
|\Lambda_1(s)\rangle-\frac{1}{4}\sec(\theta)\tan(\theta)
|\Lambda_2(s)\rangle,\\
|\kappa_4(s)\rangle-|\lambda_4(s)\rangle&=-\frac{1}{16}\sec^3(\theta)
|\Lambda_2(s)\rangle.
\end{align}
This is a satisfying result that can be anticipated from the fact 
that the operator $\mathcal{H}$ is self-adjoint, and that vectors 
$|\lambda_j(s)\rangle$ and $|\kappa_j(s)\rangle$  should always 
be solutions to the adjoint problem.

To remove the components of $\mathcal{G}$ which have an 
exponential divergence at $z=\pm\infty$, 
we impose the conditions 
\begin{align}
|\lambda_4(s)\rangle&=|0\rangle,\\
a_4|\kappa_2(s)\rangle+
b_4|\kappa_3(s)&\rangle+
c_4|\kappa_4(s)\rangle=|0\rangle.
\end{align}
Enforcing the symmetry of the Green's matrix 
$\mathcal{G}(z,s)=\mathcal{G}^\dagger(s,z)$ results in
\begin{align}
\kappa_3^4&=0,\\
\lambda_1^4&=0,\\
\kappa_3^1&=\lambda_1^3,\\
\kappa_3^2+\frac{b_4}{a_4}\kappa_3^3&=
-\frac{1}{4}\sec(\theta)\tan(\theta)\\
-\lambda_1^2&=\frac{b_4}{a_4}\kappa_3^1,
\end{align}
where $|\lambda_j(s)\rangle=\sum_{q=1}^4\lambda_j^q|\Lambda_q(s)\rangle$ and 
$|\kappa_j(s)\rangle=\sum_{q=1}^4\kappa_j^q|\Lambda_q(s)\rangle$.
% \begin{equation}
% |\Lambda_1(z)\rangle\langle\lambda_1(s)|+
% |\Lambda_2(z)\rangle\langle\lambda_2(s)|+
% |\Lambda_3(z)\rangle\langle\lambda_3(s)|=
% %
% |\kappa_1(z)\rangle\langle\Lambda_1(s)|+
% |\kappa_2(z)\rangle\langle\Lambda_2(s)|+
% |\kappa_3(z)\rangle\langle\Lambda_3(s)|+
% |\kappa_4(z)\rangle\langle\Lambda_4(s)|
% \end{equation}

The final condition to ensure the Green's matrix is unique relates to 
how the system responds to the broken symmetry caused by the 
moving impurity. The force could be either positive or negative, and the 
physically correct solution is when the force is positive 
as this is the only energy conserving solution. This gives
\begin{align}
\lambda_1^1&=0,\\
\lambda_3^3&=0,\\
\lambda_1^2&=0,
\end{align}
and completes the solution to Eq. (11) of the main article.

\begin{figure}
\includegraphics[width=0.6\textwidth]{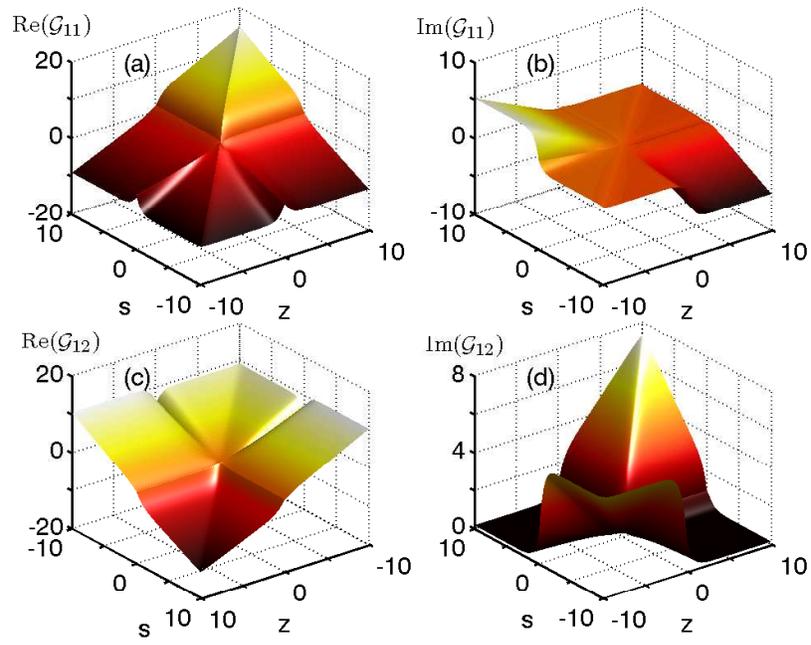}
\caption{(color online). Shows the Green's matrix 
for $\bar{\eta}=1$ and $\bar{v}=0.3$. 
(a) and (b) show the real and imaginary components respectively 
of $\mathcal{G}_{11}(z,s)$. (c) and (d) show the real and imaginary 
components respectively of $\mathcal{G}_{12}(z,s)$.}
\label{fig:green}
\end{figure}

\end{widetext}

\end{document}